\documentclass[final]{revtex4}

\usepackage[dvipdfmx]{graphicx}
\usepackage{amsmath,amssymb}
\usepackage{bm}
\usepackage{here}
\usepackage{hyperref}

\begin{document}

%=============================================
\title{Chiral Pattern in Nonrotating Spherical Convection}
%=============================================

\author{Akira Kageyama}
  \email{kage@port.kobe-u.ac.jp}
  \affiliation{Department of Computational Science, Kobe University, Kobe 657-8501, Japan}
\author{Nobuaki Ohno}
  \email{ohno@sim.u-hyogo.ac.jp}
  \affiliation{Graduate School of Information Science, University of Hyogo, Kobe 651-2197, Japan}

\begin{abstract}
When the Rayleigh number is low, Rayleigh-B\'enard convection in a nonrotating spherical shell with central gravity has symmetric solutions in terms of three-dimensional discrete rotation. All the known patterns with the regular polyhedral symmetries accompany reflection symmetry. We found a new type of steady convection in a nonrotating spherical shell by computer simulations. The pattern has the discrete rotational symmetry of a regular tetrahedron with no reflection symmetry. The convection consists of six pairs of spiral rolls placed on 12 faces of a spherical dodecahedron. Depending on the winding of the spirals, there are two possible configurations that are mirror images of one another.
\end{abstract}

\maketitle

%=============================================
\section{Introduction}

Rayleigh-B\'enard convection in a nonrotating spherical layer~\cite{Chandrasekhar1953,Chandrasekhar1953a,Busse1975} provides a venue of pattern formation that is finite but free of borders.
%Because of isotropy, convection in the low Rayleigh number regime is expected to have solutions with rotational symmetry.
In order to classify the spherical patterns,
here we refer to the five finite subgroups of $SO(3)$~\cite{Weyl1952} as %~\cite{Coxeter1972,Ihrig1984} as 
the cyclic group ($C$), 
dihedral group ($D$), 
tetrahedral group ($T$),
octahedral group ($O$), 
and icosahedral group ($I$).
Among them, we are interested in the regular polyhedral groups, $T$, $O$, and $I$.
Solutions are known for all of them;
$T$~\cite{Busse1982,Bercovici1989},
$O$~\cite{Busse1975,Bercovici1989,Futterer2010,Feudel2011},
and $I$~\cite{Busse1975,Arrial2014}.

The pattern formation on spherical surfaces in general as a bifurcation problem has been studied in detail~\cite{Matthews2003}.
Expanding a physical variable on the surface by the spherical harmonics of degree $\ell$, $Y_\ell^m$, for $-\ell\le m\le \ell$,
the problem is described by a system of equations of order of $2\ell+1$.
The degree of freedom of the system, $D(\ell)$, is uniquely determined for each subgroup of the assumed symmetry by the trace formula~\cite{Hoyle2006}.
When $D(\ell)=1$, Equivariant Bifurcation Lemma~\cite{Chossat2000} tells that there exists a unique equilibrium solution branch.
Branches and their stabilities for even $\ell$ up to 12, including $D(\ell)\ge 2$ cases, are investigated by Matthews~\cite{Matthews2003}.

In this paper, we focus on the reflection symmetry, another symmetry of a nonrotating sphere.
In theoretical models dealing with spherical pattern formation based on the reaction-diffusion equation,   % Zykov1997 Yagisita1998, 
chiral solutions naturally appear as spiral patterns on spheres~\cite{Turing1952,Varea1999,Sigrist2011,Sanchez2019}.
Also in the spherical shell convection system, a chiral solution with spiral pattern was found~\cite{Zhang2002a,Itano2015}.
This solution is a single-arm spiral in which a long roll covers the entire spherical shell with both ends on antipodal points.
It can be right- or left-handed winding.
The solution has dihedral symmetry $D$, that is, it is symmetric when flipped around an axis perpendicular to the diameter through the antipodal points~\cite{Sigrist2011}.
In the planar convection system, multiple spiral patterns, called spiral defect chaos~\cite{Morris1993,Xi1993,Liu1996}, are known to appear, but they are not chiral because the numbers of opposite windings are the same when no rotation is externally applied~\cite{Ecke1995}.

We report in this paper a class of steady, chiral, and symmetric solutions in a nonrotating spherical shell.
It has the discrete rotational symmetry of group $T$.

%Oregonator~\cite{Yagisita1998},
%Swift-Hohenberg~\cite{Matthews2003,Sigrist2011},
%FitzHugh-Nagumo~\cite{Amdjadi2005},
%and other models~\cite{Lacitignola2019}.  % a morpho-electrochemical reaction-diffusion model for alloy electrodeposition 
%a sphere is studied for buckling and folding of spherical layers~\cite{Stoop2015}.
%Axisymmetric multiple rolls in spherical shell corresponds to ``target'' pattern in planer convection~\cite{Assenheimer1993}.
%The transition from the target pattern to the spiral pattern is observed~\cite{Assenheimer1994} in the planer convection.

%=============================================
\section{Basic Equations and Method}
%=============================================

%In Rayleigh-B\'enard convections, stability of a nonlinear solution does not necessarily mean realizability of the solution.
%To demonstrate the realizability~\cite{Newell1969,Getling1998a} as well as the stability of the solutions,

We numerically integrated the time development of the fluid equation until we obtained steady solutions.
This approach was challenging in terms of both computation and visualization, 
because multiple simulation jobs with different parameters, each requiring sufficiently long integration in a diffusion timescale, have to be executed.

We assumed that an ideal gas is confined in a spherical shell layer between inner and outer spheres of radii $r_\mathrm{i}$ and $r_\mathrm{o}$, respectively.
We normalized the length by $r_\mathrm{o}=1.0$ and set $r_\mathrm{i}=0.945$.
The shell depth is $d=r_\mathrm{o}-r_\mathrm{i}=0.055$.

We used the mass density $\rho$, mass flux density $\bm{f}$, and pressure $p$ as basic variables and the flow velocity $\bm{v}=\bm{f}/\rho$ and normalized temperature $T=p/\rho$ as subsidiary variables.
We solved the time development of the variables by using the following equations.
%==========
\begin{align} 
\frac{\partial\rho}{\partial t} &=-\nabla\cdot\bm{f},  \label{210501121801}\\ 
\frac{\partial\bm{f}}{\partial t} &=-\nabla\cdot(\bm{v}\bm{f})-\nabla p - \rho\bm{g}+\mu\left\{\nabla^2\bm{v}+(1/3)\nabla(\nabla\cdot\bm{v})\right\},  \label{210501121829}\\
\frac{\partial p}{\partial t}   &= -\bm{v}\cdot\nabla p - \gamma p\nabla\cdot\bm{v}+(\gamma-1)\kappa\nabla^2 T+(\gamma-1)\Phi, \label{210501121836}
\end{align} 
%==========
where $\mu$ and $\kappa$ were dynamic viscosity and thermal diffusivity, and we assumed $\mu = \kappa$.
$\Phi$ was the dissipation function; $\Phi=2\mu \left\{ \mathrm{tr}(\epsilon \epsilon) -(1/3) \mathrm{tr}(\epsilon)^2 \right\}$ with the strain-rate tensor $\epsilon$.
Ignoring the self-gravity of the fluid, the gravity acceleration was given by $\bm{g} = G \hat{\bm{r}}/ r^2$, where $\hat{\bm{r}}$ was the unit vector in the radial direction and $G$ was a constant.
The specific heat ratio was $\gamma=5/3$.

% Boundary condition
We used rigid boundary conditions for velocity, $\bm{v}=0$, and fixed temperature conditions on $r=r_\mathrm{i}$ and $r_\mathrm{o}$.
% Initial condition
The initial temperature profile $T_0(r)$, pressure $p_0(r)$, and density $\rho_0(r)$ were spherically symmetric hydrostatic equilibrium states with thermal conduction.
The initial velocity was $\bm{v}_0=0$.
We normalized the thermodynamic variables in terms of their values on $r=r_\mathrm{o}(=1)$,
that is, $T_0(r_\mathrm{o})=p_0(r_\mathrm{o})=\rho_0(r_\mathrm{o})=1.0$.
The initial state was given as
$T_0(r) = \beta/r + 1 - \beta$, 
$p_0(r) = T_0^n$,
and $\rho_0(r) = T_0^{n-1}$.
Here, $\beta$ was set such that $T_0(r_\mathrm{i}) = 1.05$; further, $n=1.25$ and $G=n\beta$.
In the simulations, the Mach number of the flow was at most $3.7\times 10^{-3}$.
The density stratification was also small, being $\rho(r_\mathrm{i})\sim 1.012$. 
Time was normalized by the diffusion timescale $\tau_\mathrm{d}=d^2/\mu=d^2/\kappa$.
%To present flow amplitude, we use free fall velocity defined by $v_\mathrm{ff} = \sqrt{\delta TG(1/r_\mathrm{i}-1})$
%with $\delta T= T_0(r_\mathrm{i})-1$ as the unit of velocity.

% Definition of Rayleigh number
The Rayleigh number of an ideal gas depends on the radius~\cite{Gilman1981} as
%==========
\begin{equation}  \label{210501210209} 
  Ra = \frac{g\left\{-(dT/dr)/T-g/c_p T\right\}d^4}{\left(\kappa/c_p \rho\right)\left(\mu/\rho\right)}
= \frac{n \beta^2  (c_p-n) d^4}{\kappa \mu }f(r).
\end{equation} 
%==========
where $c_p=\gamma/(\gamma-1)$ is the specific heat.
The radial factor $f(r)=T(r)^{2n-3}/r^4$ does not change greatly in the shell; $f(r_\mathrm{o})=1$
and $f(r_\mathrm{i}) \sim 1.224$. % $f(\bar{r}) = 1.10466$. 
Below, we estimated $Ra$ at the middle of the shell~\cite{Spiegel1965} at $\bar{r} = (r_\mathrm{o}+r_\mathrm{i})/2$ and denoted it as $\bar{R}$.

%---<in the first submission>---
% Discretization 
%We discretized the right-hand sides of Eqs.~\eqref{210501121801}--\eqref{210501121836} 
%using a second-order finite difference method on a Yin-Yang grid~\cite{Kageyama2004}, in which two congruent component grids, the Yin grid and the Yang grid, are combined in a complementary way to cover a spherical shell based on the standard overset grid methodology.
%The total grid size was $N_r\times N_\vartheta\times N_\varphi\times 2 = 60 \times 404 \times 1208 \times 2$
%in the radial ($r$), colatitudinal ($\vartheta$), and longitudinal ($\varphi$) directions.
%The last factor 2 was for the Yin and Yang components.
%For numerical integration over time, we used a fourth-order explicit Runge--Kutta method.
%---<in the first submission>---
%
%---<in the 2nd submission>---
% Discretization 
We discretized the right-hand sides of Eqs.~\eqref{210501121801}--\eqref{210501121836} 
using a second-order finite difference method on a Yin-Yang grid~\cite{Kageyama2004}.
The total grid size was $N_r\times N_\vartheta\times N_\varphi\times 2 = 60 \times 404 \times 1208 \times 2$
in the radial ($r$), colatitudinal ($\vartheta$), and longitudinal ($\varphi$) directions.
The last factor 2 was for the Yin and Yang components.
For numerical integration over time, we used a fourth-order explicit Runge--Kutta method.

We used VISMO-YY~\cite{Ohno2021b} to generate in-situ visualization images of the isosurfaces of radial velocity, that is, $v_\mathrm{r}(r,\vartheta,\varphi)=\pm \alpha$, where $\alpha$ is constant.
VISMO-YY is a parallelized open-source software rendering library that is specialized for the in-situ visualization of Yin-Yang grid simulations~\cite{Ohno}.
The level of the isosurface $\alpha$ was set to half the maximum radial velocity, $\alpha = 0.5\times \text{max}\{|v_\mathrm{r}|\}$, at each snapshot.
The rising fluid $(v_\mathrm{r}=+\alpha)$ and sinking fluid $(v_\mathrm{r}=-\alpha)$ are respectively colored in orange and blue in the following images.

To apply the idea of Four-Dimensional Street View (4DSV)~\cite{Kageyama2020}, that is, in-situ visualization with multiple viewpoints, 
we placed 10 visualization cameras, $C_1, C_2, \ldots, C_{10}$, around the spherical shell.
The distance from the origin to the cameras was $r=2.6$.
Cameras $C_k \ (1\le k\le 8)$ were placed on the ``equator,'' that is, at $\vartheta=\pi/2$ and $\varphi=(k-1)\pi/4$.
Cameras $C_9$ and $C_{10}$ were located above the north and south poles, that is, at $\vartheta=0$ and $\pi$, respectively.
% %===<Removing in the second submit>===
% The cameras synchronously took $512\times 512$ pixel snapshots for visualization.
% Around $20\%$ of the total simulation time of each job was spent on the in-situ visualization by the cameras.
% %===</Removing in the second submit>===

%=============================================
\section{Critical Rayleigh Number}
%=============================================

%---<new paragraph in the 2nd submit>---
A standard approach to spherical pattern formation is the stability analysis for each degree $\ell$ of the spherical harmonics $Y_\ell^m$. 
Zhang et al.~\cite{Zhang2002a}~performed linear computations for convection in a thin spherical shell with radii ratio $r_i/r_o\sim 0.8475$ and found that the critical Rayleigh numbers $R_\mathrm{c}$ for $\ell=17, 18, 19$, and $20$ are $1723.5$, $1710.7$, $1712.4$, and $1727.1$, respectively. 
The $R_\mathrm{c}$ values are very close to each other. 
Because the spherical shell in our study ($r_i/r_o=0.945$) is thinner than theirs, there would be more modes (having larger degree $\ell$) with close $R_\mathrm{c}$ values. 
The mixed existence of multiple unstable modes with large $\ell$ values means that it is highly demanding to investigate the nature of the bifurcation for the onset of convection in our thin spherical shell. 
In addition to that, our numerical method, which is a point-based spatial discretization rather than the commonly used spectral method using spherical harmonics, makes the numerical stability analysis hard for each degree $\ell$.
%---</new paragraph in the 2nd submit>---

% Critical Rayleigh number (Random perturbation)
To find the critical Rayleigh number of this system, 
we performed parameter runs with different Rayleigh numbers $\bar{R}$ under random perturbations on the pressure in the initial condition as follows:
%==========
\begin{equation}  \label{210502174420} 
  p(r,\vartheta,\varphi) = p_0(r)  + c\, p_1(\vartheta,\varphi)\, \sin{\left\{\pi (r-r_\mathrm{i})/d\right\}},
\end{equation} 
%==========
where $c$ is a small positive number.
The perturbation profile $p_1$ is a linear combination of spherical harmonics.
%==========
\begin{equation}  \label{210410113708} 
  p_1(\vartheta,\varphi) 
     = \sum_{\ell=1}^{L_\mathrm{max}} \sum_{m=0}^\ell 
        \delta_\ell^m \, 
        \hat{P}_\ell^m(\vartheta,\varphi) \cos{(m\varphi+d_{\ell}^m)},
\end{equation}
%==========
where $L_\mathrm{max}=202$;
$\hat{P}_\ell^m$  are normalized Legendre function;
$\delta_\ell^m$ are binary coefficient ($\delta_\ell^m=0$ or $1$);
and $d_\ell^m$ are random phase between 0 and $2\pi$.
We randomly selected $5\%$ of all possible $(\ell, m)$ pairs and set $\delta_\ell^m=1$ for them
and $\delta_\ell^m=0$ for the others.
The constant $c$ in eq.~\eqref{210502174420} was specified after these random pickups so that the maximum amplitude of the resulting perturbation all over the shell, that is, $\text{max}\{|c\, p_1(\vartheta,\varphi)|\}$, was $1.0\times 10^{-3}$.

When we changed $\bar{R}$, we fixed $\beta$ and other constants in the numerator of the last term in eq.~\eqref{210501210209}.
In other words, we changed only the diffusivities $\mu (=\kappa)$ in these parameter runs.

%-------------------------------------------------------
\begin{figure}[H]   \centering   
  \includegraphics[%
     height=0.6\textheight,%
       width=0.6\hsize,keepaspectratio]%
         {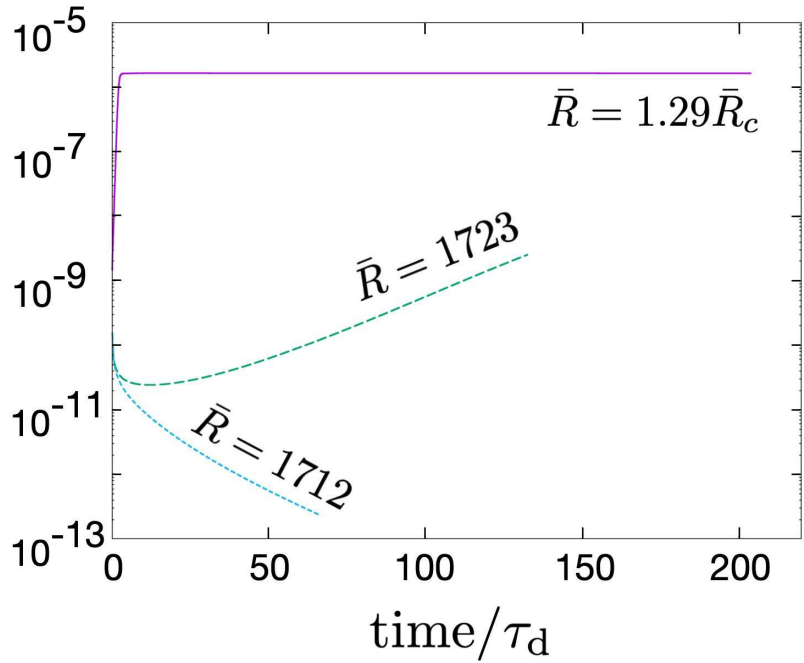}   
      \caption{Time developments of convection flow energy for three different Rayleigh numbers $\bar{R}$.
      Time was normalized by the diffusion time $\tau_\mathrm{d}$.
      The dotted blue curve ($\bar{R}=1712$) and dashed green curve ($\bar{R}=1723$) show simulation results obtained starting from random perturbations.
      We assumed that the critical Rayleigh number $\bar{R}_c$ was the midpoint of the two $\bar{R}$ values.
      The solid magenta curve ($\bar{R}=2209=1.29\bar{R}_c$) shows the simulation result obtained starting from a controlled initial condition.}
      \label{210510170600}
\end{figure} 
%-------------------------------------------------------

We performed several jobs to find the critical $\bar{R}$; 
Fig.~\ref{210510170600} shows two decisive runs among these.
In this figure, the horizontal axis represents the simulation time normalized by the diffusion time $\tau_\mathrm{d}$,
and the vertical axis represents the total energy of the convection flow.
We first focus on the case of $\bar{R}=1712$ (blue dotted line).
The flow initiated by the random perturbation decays, or the fluid is stable.
By contrast, in the case of $\bar{R}=1723$ (green dashed line), the energy increases exponentially with time.
These observations indicate that the critical Rayleigh number $\bar{R}_\mathrm{c}$ is between these two values.
Here, we assume that it is the midpoint, that is, $\bar{R}_\mathrm{c}=(1712+1723)/2=1717.5$, which is close to the value of 1708 for convection in horizontal planes~\cite{Chandrasekhar1981a}.
It is known that $\bar{R}_\mathrm{c}$ in spherical thin shells becomes slightly larger than that in the plane layer convection~\cite{Zhang2002a}.

%-------------------------------------------------------
\begin{figure}[H]   \centering   
  \includegraphics[%
     height=0.6\textheight,%
       width=0.8\hsize,keepaspectratio]%
         {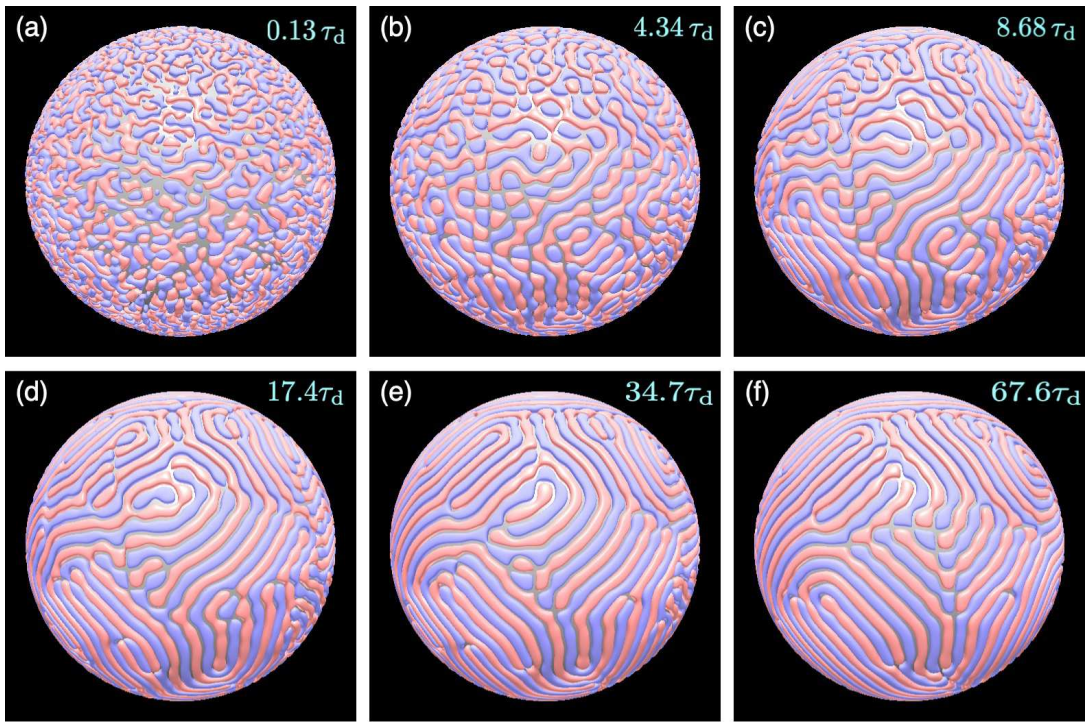}   
      \caption{Time development of convection for $\bar{R}=1.29\bar{R}_\mathrm{c}$ when initial condition is random.
      Time is presented on the diffusion timescale $\tau_\mathrm{d}$ in the upper-right corner of each panel.
      From~(d) to~(f), the pattern slowly changes in time.}
      \label{211016160509}
\end{figure} 
%-------------------------------------------------------

% ===<new paragraph for 2nd submit>===
Fig.~\ref{211016160509} shows snapshots of convection when $\bar{R}=1.29\bar{R}_\mathrm{c}$ % Ra=2000
started from the random initial condition used to determine the critical Rayleigh number.
The random cellular pattern observed in the very early stage in Figs.~\ref{211016160509}(a) and~(b) is replaced by a set of unsteady rolls as shown in Figs.~\ref{211016160509}(c) to~(f).
It resembles the onset of spiral defect chaos in the plane layer convection at small Prandtl number~\cite{Morris1993,Xi1993,Liu1996}.
The pattern is constantly moving and there is no sign of settling down to a steady state.
% ===</new paragraph for 2nd submit>===

%
%Fig.~\ref{210408172355} shows snapshots of the convection when $\bar{R}=1723$ visualized by isosurfaces of radial component of velocity $v_r$.
%Panels~(a) and~(b) show transient phase at $t= 0.001\, \tau_\mathrm{d}$, and
%$t=1.36\, \tau_\mathrm{d}$, respectively.
%The convection pattern almost settles down in panel~(c), although the flow energy is still slowly growing.
%The maximum velocity at (c) $t=132.5\tau_\mathrm{d}$  % (c)
%is slow; 
%$v_\mathrm{max} = 7.41\times 10^{-3}\, v_\mathrm{ff}$
%				                     % vmax =4.1404245167721304/10000  in simulation unit
%				                     %         = .00740661654016736090 * v_\mathrm{ff}

%=============================================
\section{Initial Condition}
%=============================================

%Initial condition determines the finite-amplitude behavior of the fluid~\cite{Young1974}.

For the numerical demonstration of steady chiral convection,
we adopted the controlled initial condition method in convection experiments~\cite{Chen1968,Busse1974}.
Our strategy was to specify a perturbation in the initial condition with chirality and find if it led to a steady solution without breaking the chirality.

\begin{figure}[H]   \centering   
  \includegraphics[%
     height=0.6\textheight,%
       width=0.6\hsize,keepaspectratio]%
         {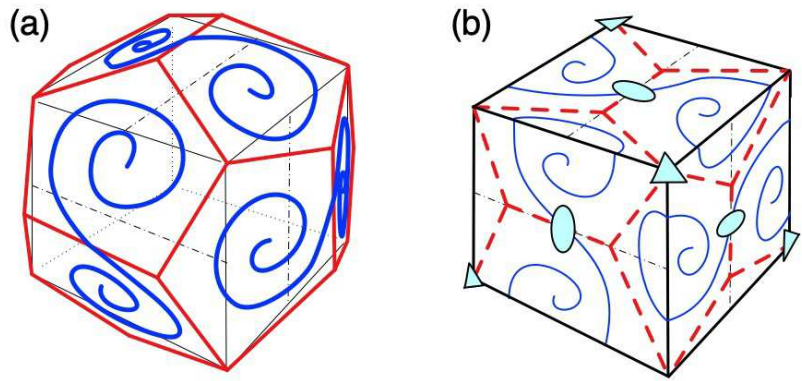}   
      \caption{Design of initial condition with chirality.
      (a) Six pairs of Archimedean spirals (blue) were placed on 12 faces of a regular dodecahedron (red).
      (b) Discrete rotational symmetry of $T$ (tetrahedral) group, 
       with four 3-fold rotations indicated by light blue triangles and three 2-fold rotations indicated by light blue ellipses.
      }
      \label{210411165738}
\end{figure}

For the initial condition, we constructed 12 Archimedean spirals with the same winding [blue curves in Fig.~\ref{210411165738}(a)] on a regular dodecahedron [red lines in Fig.~\ref{210411165738}(a)].
Next, we selected six adjacent pairs from the 12 faces of the dodecahedron and smoothly connected two spirals on each pentagon together.
We called a pair of connected spirals a dipole~\cite{Ecke1995}.

Among the various possible combinations of the six pairs of pentagonal faces,
we selected the special one shown in Fig.~\ref{210411165738}(a) that has a discrete rotational symmetry.
A dodecahedron can be constructed by adding a ``roof'' to each face of a cube~\cite{Cromwell1997}.
The rotational symmetry of the dipoles is much easier to understand if we observe the underlying cube with a spiral texture,
as shown in Fig.~\ref{210411165738}(b) 
%-------------------------------------------------------
%-------------------------------------------------------

The rotational symmetries of a plain cube partially disappear owing to the spiral texture.
The remaining rotational symmetries of the texture shown in Fig.~\ref{210411165738}(b) are
four 3-fold rotations with axes about the cube's diagonal [indicated by light blue triangles in Fig.~\ref{210411165738}(b)]
and three 2-fold rotations about axes perpendicular to square faces [indicated by light blue ellipses in Fig.~\ref{210411165738}(b)].
The dodecahedron with the spiral texture has the discrete rotational symmetry of the tetrahedral group $T$~\cite{Cromwell1997}.

\begin{figure}[H]   \centering   
  \includegraphics[%
     height=0.6\textheight,%
       width=0.6\hsize,keepaspectratio]%
         {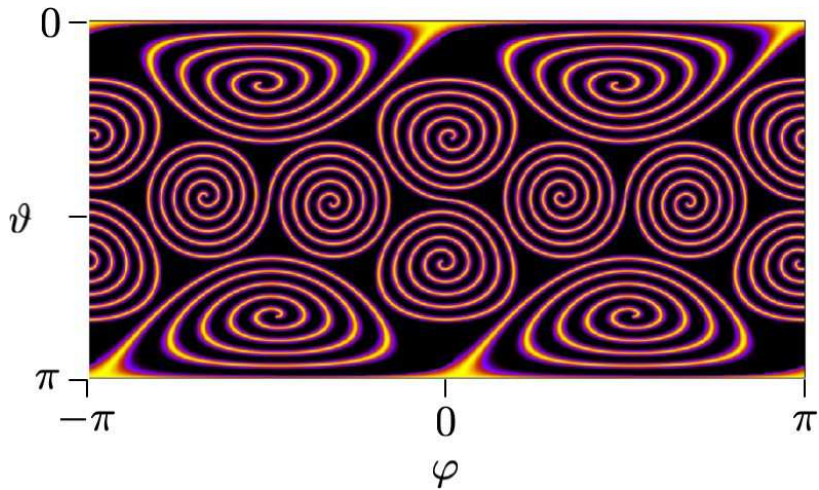}   
      \caption{Pressure perturbation profile $p_1(\vartheta,\varphi)$ used in the controlled initial condition in this study.}
      \label{210410181714}
\end{figure}

We took the above $T$-symmetric dipole curves as the skeleton of a pressure perturbation $p_1(\vartheta,\varphi)$ in Eq.~\eqref{210502174420}.
We first mapped the dipole curves on the regular dodecahedron onto a unit sphere, and for each point $(\vartheta,\varphi)$ on the sphere, 
we found the distance $\delta$ to the closest dipole curve on the sphere from the point.
We then specified the perturbation $p_1(\vartheta,\varphi)$ in terms of a Gaussian function of $\delta$.
A free parameter in the configuration of the skeleton was the winding number $n$ of each spiral in a pentagon.
Fig.~\ref{210410181714} shows the $p_1$ profile when $n=5$, which was used in all simulations described below.
Depending on the winding of the spirals, there are two possible configurations that are mirror images of one another.
A set of programs to calculate $p_1(\vartheta,\varphi)$ and to draw Fig.~\ref{210410181714} is available at GitHub~\cite{KageyamaURL}.
As in the case of random perturbations, the constant $c$ in eq.~\eqref{210502174420} was adjusted so that $\text{max}\{|c\, p_1(\vartheta,\varphi)|\} =1.0\times 10^{-3}$.

%-------------------------------------------------------
%-------------------------------------------------------

%=============================================
\section{Chiral Pattern Convection}
%=============================================

\begin{figure}[H]   \centering   
  \includegraphics[%
     height=0.6\textheight,%
       width=0.99\hsize,keepaspectratio]%
       {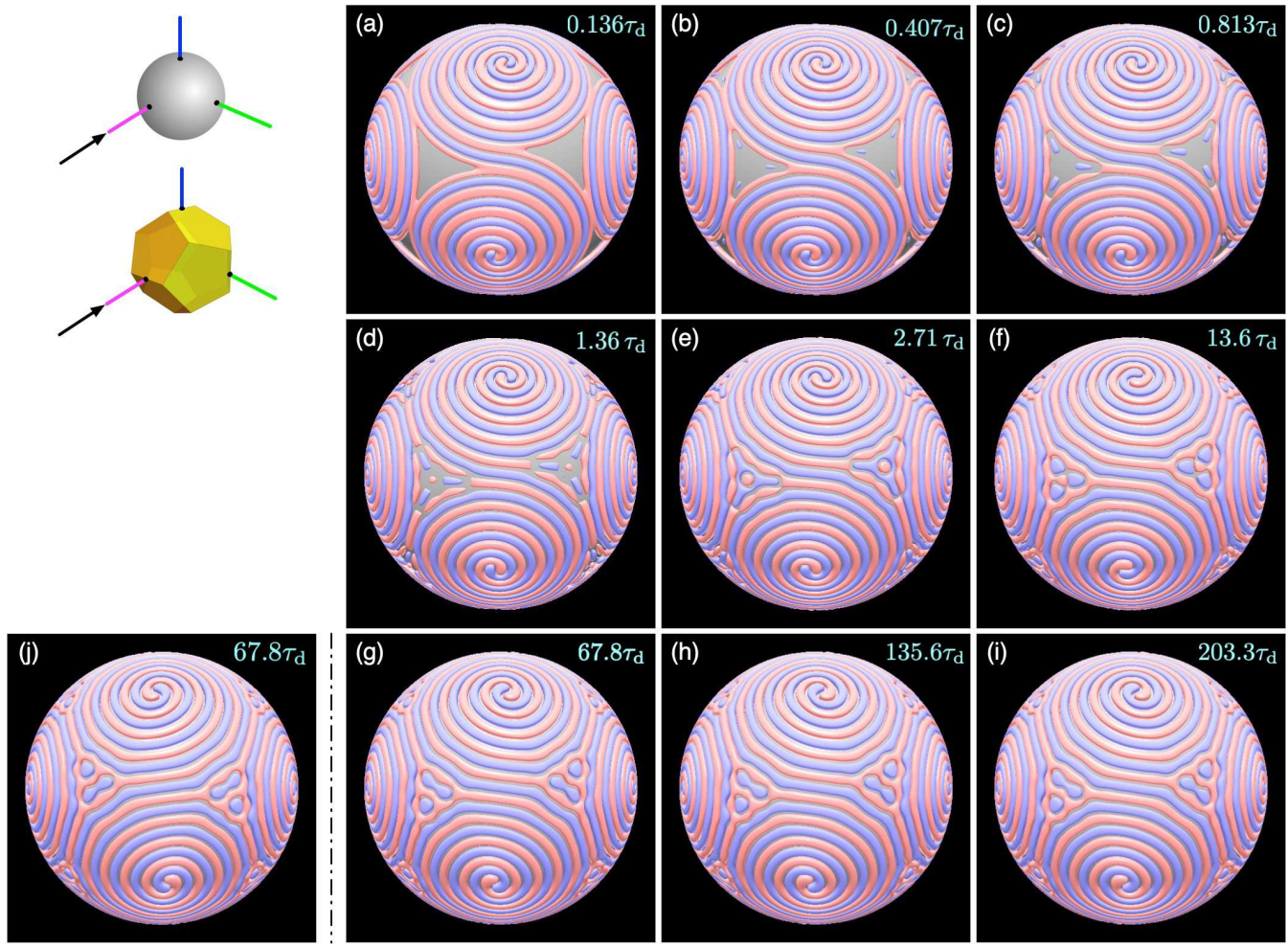}   
      \caption{Time development of convection when $\bar{R}=1.29\bar{R}_\mathrm{c}$.
      Time is presented on the diffusion timescale $\tau_\mathrm{d}$ in the upper-right corner of each panel.
      The convection reached the steady state by $t=67.8~\tau_\mathrm{d}$.
      No shift or rotation of spiral arms was observed.
      Panels (a) to~(i) show the convection started by the pressure perturbation shown in Fig.~\ref{210410181714}.
      Panel~(j) shows a snapshot of another simulation started by the mirrored perturbation of Fig.~\ref{210410181714}.}
      %===========================================
      % job: r0p945R2000DodecaM5
      % Raw data
      %     rad_min = 0.945
      %     d = 1-0.945 = 0.055
      %     max flow_velocity = 4.8070E-3
      %     time_turnover=2*(1-0.945)/4.8070E-3=22.883
      %     kappa = 7.2637270706594879E-05
      %     time_kappa = d^2/kappa = 41.6452871999960 --> tau=41.6
      %   (a) nloop=    100,000, t= 56.3970380002073810 =  1.35569802885113896634 tau
      %   (b) nloop=    200,000, t=1.1279009636242405E+02 = 2.71130039332750120192 tau
      %   (c) nloop=  1,000,000, t=5.6392655034628910E+02 = 13.55592669101656490384 tau
      %   (d) nloop=  5000,000, t=2.8195932728214557E+03 =  67.77868444282345432692 tau
      %   (e) nloop=10,000,000, t=5.6391705268171854E+03=  135.55698381772080288461 tau
      %   (f) nloop= 15,000,000, t=8.4587469131611269E+03 = 203.33526233560401201923 tau
      %
      %%   (a) nloop=4000, t=2.2558676606100585 = 0.05422 tau
      %%   (b) nloop=74000, t=41.7341477267440339 = 1.0032 tau
      %%   (c) nloop=738000, t=4.1617123949190564E+03 = 100 tau
      %%   (d) nloop=15000000, t=8.4587469131611269E+03 = 203.335 tau
      %===========================================
      \label{210410151013}
\end{figure}

We first present the simulation results for $\bar{R}=1.29\bar{R}_\mathrm{c}$. % Ra=2000
The chiral profile similar to the initial perturbation appeared in the convection flow, which becomes steady. 
Fig.~\ref{210410151013} shows a sequence of snapshots taken by camera $C_1$ 
from $t=0.136\,\tau_\mathrm{d}$ 
%from $t=1.36\,\tau_\mathrm{d}$ % nloop=4000  <== 1st submit
to $t=203.3\, \tau_\mathrm{d}$. % nloop=15000000
The viewing direction of $C_1$ is shown by the arrow in the diagrams in the upper-left,
in which the red, green, and blue lines denote $x$, $y$, and $z$ axes, respectively, and the yellow solid below the gray sphere represents configuration of the regular dodecahedron used in the pressure perturbation.
%As shown below, the convection kept rotational $T$-symmetry.

The convection appeared according to the initial stripe pattern of the pressure perturbation
as shown in Fig.~\ref{210410151013}(a) at $t=0.136~\tau_\mathrm{d}$.
Each pentagon's spiral slowly rotated around its axis. 
In the initial phase, a more notable change was observed at the five corners around each pentagon, which correspond to the vertices of the original regular dodecahedron.
The flow was initially absent at the corner regions before they were gradually filled with convection.
The filling pattern in each corner seemed to have three-fold rotational symmetry in the early stage,
as indicated by the three short blue bars around the small red point in Fig.~\ref{210410151013}(d). 
The three-fold rotational symmetry is still observed in Fig.~\ref{210410151013}(e) ($t=2.71~\tau_\mathrm{d}$), following which the corner slowly loses its three-fold symmetry (see Fig.~\ref{210410151013}(f), $t=13.6~\tau_\mathrm{d}$).
The convection reached almost the final pattern by $t=13.6\, \tau_\mathrm{d}$,
although a slight adjustment of the whole pattern was still observed.
%-------------------------------------------------------
%-------------------------------------------------------

%Steady state is realized by the time of $t=67.8~\tau_\mathrm{d}$ 
Figs.~\ref{210410151013}(g)--(i) show the convection in the last two-thirds of the simulation time.
The three panels appeared identical, which was true in views from other cameras.
This indicated that the convection had reached a steady state.
No rotation of the arms was observed, unlike the global or local spirals in the case of planar convections~\cite{Bodenschatz1991a,Vitral2020}.
The same was true of other steady solutions with different $\bar{R}$, as described below.

The twelve spirals in the steady state had the same winding direction as the initial pattern of the pressure perturbation.
We can expect that mirrored pattern with the opposite winding is also stable.
We confirmed this by performing a simulation with the initial condition of pressure perturbation that is the mirror image of Fig.~\ref{210410181714}.
The time development as well as the final steady state of the convection was the same, except that they had the opposite winding.
A snapshot in the final steady state is shown in Fig.~\ref{210410151013}(j) in the lower-left of the figure.
The two final states shown in Fig.~\ref{210410151013}(j) and in Fig.~\ref{210410151013}(g) [or panels (h) and (i)] are mirror images one another.

\begin{figure}[H]   \centering   
  \includegraphics[%
     height=0.6\textheight,%
       width=0.8\hsize,keepaspectratio]%
         {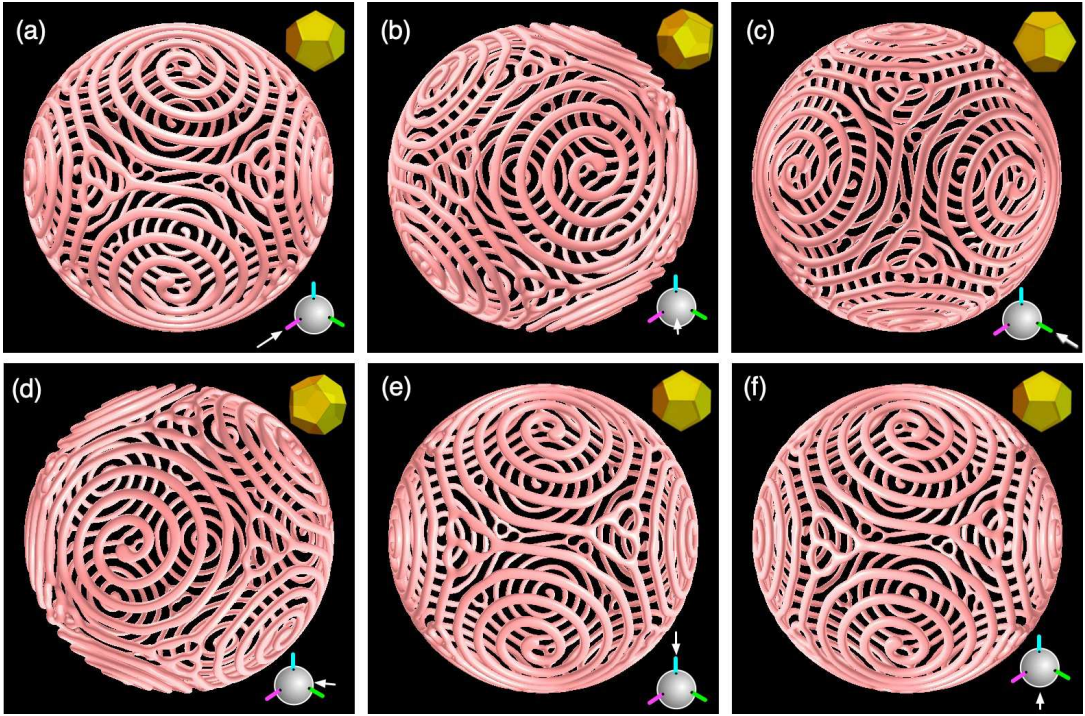}   
      \caption{Dodecahedral convection pattern with discrete rotational symmetry of $T$ for $\bar{R}=1.29 \bar{R}_\mathrm{c}$.
      Isosurfaces of positive $v_r$ were visualized using six different in-situ visualization cameras.
      The viewing direction from each camera was indicated by arrows in the lower right.}
      \label{210408181558}
\end{figure}

Fig.~\ref{210408181558} shows the isosurfaces of positive $v_r$ in the steady state 
when $\bar{R}=1.29\bar{R}_\mathrm{c}$. % Ra=2000
Snapshots for in-situ visualization were taken at the same time 
($t=203.3\, \tau_\mathrm{d}$) %(nloop=015100000)
from six different visualization cameras: $C_1$--$C_4$ for (a)--(d)
and $C_9$ and $C_{10}$ for (e) and~(f).
The viewing direction of each camera is indicated by the arrow in the lower right in each panel.
The spirals were located on the faces of a spherical dodecahedron, and the whole pattern kept rotational $T$-symmetry with the chirality of the initial perturbation.
%-------------------------------------------------------
%-------------------------------------------------------

\begin{figure}[H]   \centering   
  \includegraphics[%
     height=0.6\textheight,%
       width=0.8\hsize,keepaspectratio]%
         {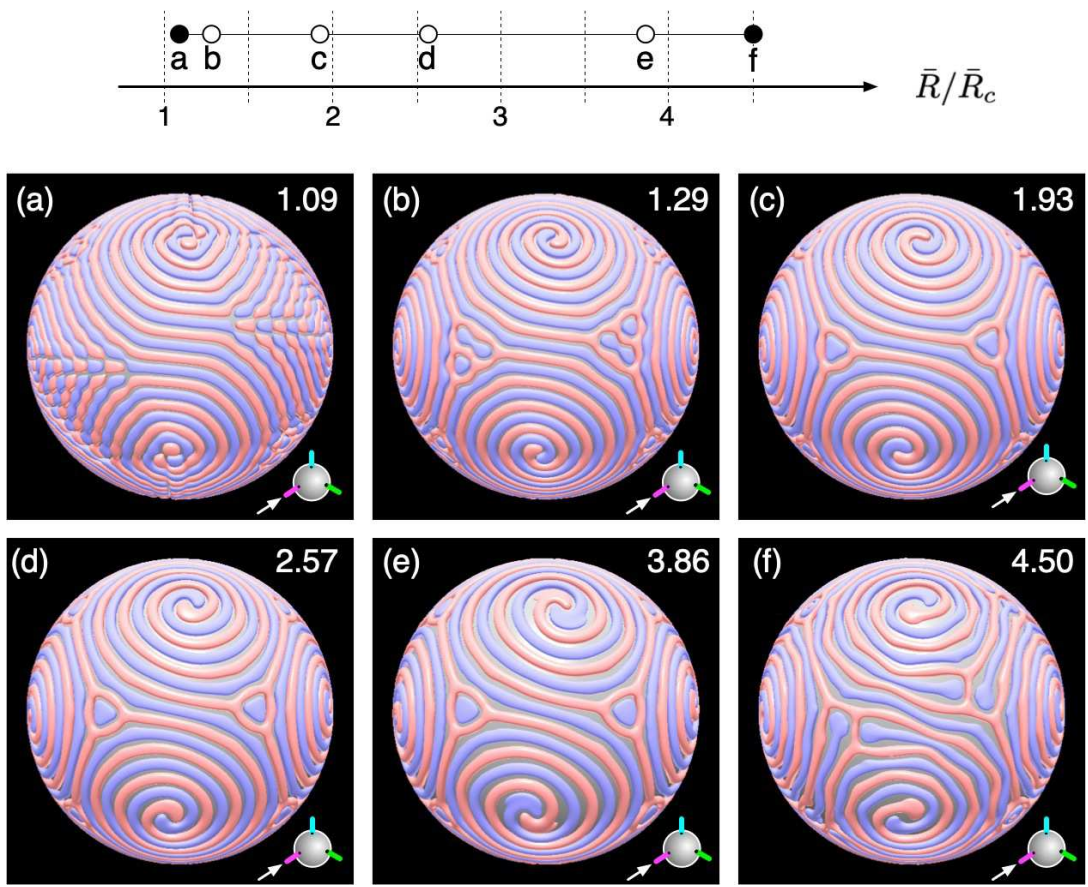}   
      \caption{%
      Black or white circles with labels \texttt{a}--\texttt{f} in the upper part show the range of Rayleigh numbers $\bar{R}$ for steady-state solutions;
      white circles indicate that steady, chiral, and symmetric solutions are obtained by simulations.
      The convection patterns are presented in the lower six panels with corresponding labels.
      Snapshots were taken at $t=113~\tau_\mathrm{d}$ for (a)--(e), and at $t=12.2~\tau_\mathrm{d}$ for~(f).%
      }
%      A number in the upper-right corner of each panel is normalized value $\bar{R}/\bar{R}_\mathrm{c}$.
%      Stationary solutions with $T$ symmetry are shown in white circles in the diagram in the lower-right.
%      Panels (a) and~(f) show convections for non-steady, non-symmetric solutions.
%      Panels (b) to~(e) are steady, chiral, and symmetric solutions.}
      % (a) Ra=1700 (Ra/Ra_c=1.0932475884)  nloop=8300000, time=4.680620E+03, camera position = 01
      % (b) Ra=2000
      % (c) Ra=3000
      % (d) Ra=4000
      % (e) Ra=6000
      % (f) Ra=7000 (Ra/Ra_c=4.501607717) nloop=900000, time = 5.07517E+02,  camera position = 01
      \label{210412185344}
\end{figure}

Starting from the same $p_1$ perturbation,
we also performed simulations with different $\bar{R}$ values.
The results are summarized in the upper part in Fig.~\ref{210412185344}.
%The range of $\bar{R}$ for steady solutions is shown.
Black circles (a and f) correspond to unsteady, while white circles (b to e) correspond to steady convection.
The steady solutions span a range of Rayleigh numbers,
$1.29\bar{R}_\mathrm{c} \le \bar{R} \le 3.86\bar{R}_\mathrm{c}$.
They have basically the same pattern, with slight variations in the pentagonal corners.

The lower six panels in Fig.~\ref{210412185344} show convection patterns for different $\bar{R}$.
All panels except Fig.~\ref{210412185344}(f) show snapshots taken at $t=113~\tau_\mathrm{d}$.
%which corresponds more than $8,400,000$ simulation steps. % time =    4.7219329729405272E+03
%The panels (a) and~(f) are located outside the range for steady state.

We first examine Fig.~\ref{210412185344}(a) for $\bar{R}=1.09\bar{R}_\mathrm{c}$.
The $T$-symmetric spirals on the regular dodecahedron, similar to the case of $\bar{R}=1.29\bar{R}_\mathrm{c}$, grew initially but were just barely unstable. 
Ripple-like modes along the spiral arms appeared in Fig.~\ref{210412185344}(a).
As the ripple grew, the rolls were broken and the convection pattern shifted to irregular cells.

Fig.~\ref{210412185344}(b) shows the steady solution for $\bar{R}=1.29\bar{R}_\mathrm{c}$, that we have examined above.
Fig.~\ref{210412185344}(c)--(e) also show solutions in the steady solution range.

The convection pattern under $\bar{R}=4.50\bar{R}_c$ was unstable, as shown in Fig.~\ref{210412185344}(f).
The collapse of the pattern was so fast that we show a snapshot at $t=12.2~\tau_\mathrm{d}$ in this case.  % time = 5.07517E+02

The steady states shown in panels (b)--(e) indicate that the chiral pattern of six sets of spiral dipoles on a spherical dodecahedron is a stable solution in a nonrotating spherical shell in the Rayleigh number regime.

%%-------------------------------------------------------
%\begin{figure}[H]   \centering   
%  \includegraphics[%
%     height=0.6\textheight,%
%       width=0.6\hsize,keepaspectratio]%
%         {./figs/broken_pattern_with_Ra1700_Ra7000.jpg}   
%      \caption{(a) $Ra/Ra_\mathrm{c}=1.290$.}
%      % (a) Ra=1700 (Ra/Ra_c=1.0932475884)  nloop=8300000, time=4.680620E+03, camera position = 01
%      % (b) Ra=7000 (Ra/Ra_c=4.501607717) nloop=900000, time = 5.07517E+02,  camera position = 01
%      \label{210412185344}
%\end{figure} 
%%-------------------------------------------------------

%-------------------------------------------------------
%-------------------------------------------------------

%=============================================
\section{Conclusion}
%=============================================

Because of geometrical isotropy,  
Rayleigh-B\'enard convection in a spherical vessel with central gravity is expected to exhibit symmetric patterns in terms of rotation.
All three polyhedral groups, namely, $T$ (tetrahedral), $O$ (octahedral), and $I$ (icosahedral), are known to appear in the solutions of spherical shell convection.

We found a new type of chiral solution in a nonrotating spherical shell.
The solution had discrete rotational symmetry of group $T$.
The pattern consisted of six pairs of spiral rolls placed on 12 faces of a spherical dodecahedron.
The solution kept chirality as well as rotational symmetry for Rayleigh numbers in the range of $\bar{R}=1.29~\bar{R}_\mathrm{c}$ to $3.86~\bar{R}_\mathrm{c}$.

The chiral solution was found by using a carefully adjusted initial condition.
Chiral patterns with discrete rotational symmetry of the $O$ or $I$ group might exist; however, we have not yet explored this possibility.
\citet{Matthews2003} used the bifurcation theory and showed that $I$-symmetry is preferred in pattern formations on a sphere in general.
An $I$-symmetric chiral solution could possibly be constructed using the same kind of controlled initial condition method adopted in the present study.
%Chirality would be useful to classify solutions in nonrotating spherical shell.

%=============================================
\section*{Acknowledgments}
%=============================================
This work was supported by JSPS KAKENHI Grant Number 17H02998.
This work was performed on ``Plasma Simulator'' (NEC SX-Aurora TSUBASA) of NIFS with the support and under the auspices of the NIFS Collaboration Research program (NIFS15KNSS062).

%%\nocite{*}
%\bibliography{210412}

\begin{thebibliography}{35}%
\makeatletter
\providecommand \@ifxundefined [1]{%
 \@ifx{#1\undefined}
}%
\providecommand \@ifnum [1]{%
 \ifnum #1\expandafter \@firstoftwo
 \else \expandafter \@secondoftwo
 \fi
}%
\providecommand \@ifx [1]{%
 \ifx #1\expandafter \@firstoftwo
 \else \expandafter \@secondoftwo
 \fi
}%
\providecommand \natexlab [1]{#1}%
\providecommand \enquote  [1]{``#1''}%
\providecommand \bibnamefont  [1]{#1}%
\providecommand \bibfnamefont [1]{#1}%
\providecommand \citenamefont [1]{#1}%
\providecommand \href@noop [0]{\@secondoftwo}%
\providecommand \href [0]{\begingroup \@sanitize@url \@href}%
\providecommand \@href[1]{\@@startlink{#1}\@@href}%
\providecommand \@@href[1]{\endgroup#1\@@endlink}%
\providecommand \@sanitize@url [0]{\catcode `\\12\catcode `\$12\catcode
  `\&12\catcode `\#12\catcode `\^12\catcode `\_12\catcode `\%12\relax}%
\providecommand \@@startlink[1]{}%
\providecommand \@@endlink[0]{}%
\providecommand \url  [0]{\begingroup\@sanitize@url \@url }%
\providecommand \@url [1]{\endgroup\@href {#1}{\urlprefix }}%
\providecommand \urlprefix  [0]{URL }%
\providecommand \Eprint [0]{\href }%
\providecommand \doibase [0]{https://doi.org/}%
\providecommand \selectlanguage [0]{\@gobble}%
\providecommand \bibinfo  [0]{\@secondoftwo}%
\providecommand \bibfield  [0]{\@secondoftwo}%
\providecommand \translation [1]{[#1]}%
\providecommand \BibitemOpen [0]{}%
\providecommand \bibitemStop [0]{}%
\providecommand \bibitemNoStop [0]{.\EOS\space}%
\providecommand \EOS [0]{\spacefactor3000\relax}%
\providecommand \BibitemShut  [1]{\csname bibitem#1\endcsname}%
\let\auto@bib@innerbib\@empty
%</preamble>
\bibitem [{\citenamefont
  {Chandrasekhar}(1953{\natexlab{a}})}]{Chandrasekhar1953}%
  \BibitemOpen
  \bibfield  {author} {\bibinfo {author} {\bibfnamefont {S.}~\bibnamefont
  {Chandrasekhar}},\ }\bibfield  {title} {\bibinfo {title} {{The onset of
  convection by thermal instability in spherical shells}},\ }\href
  {https://doi.org/10.1080/14786440308520302} {\bibfield  {journal} {\bibinfo
  {journal} {Phyl. Mag.}\ }\textbf {\bibinfo {volume} {44}},\ \bibinfo {pages}
  {233} (\bibinfo {year} {1953}{\natexlab{a}})}\BibitemShut {NoStop}%
\bibitem [{\citenamefont
  {Chandrasekhar}(1953{\natexlab{b}})}]{Chandrasekhar1953a}%
  \BibitemOpen
  \bibfield  {author} {\bibinfo {author} {\bibfnamefont {S.}~\bibnamefont
  {Chandrasekhar}},\ }\bibfield  {title} {\bibinfo {title} {{The onset of
  convection by thermal instability in spherical shells (A correction)}},\
  }\href {https://doi.org/10.1080/14786441008520371} {\bibfield  {journal}
  {\bibinfo  {journal} {Phyl. Mag.}\ }\textbf {\bibinfo {volume} {44}},\
  \bibinfo {pages} {1129} (\bibinfo {year} {1953}{\natexlab{b}})}\BibitemShut
  {NoStop}%
\bibitem [{\citenamefont {Busse}(1975)}]{Busse1975}%
  \BibitemOpen
  \bibfield  {author} {\bibinfo {author} {\bibfnamefont {F.~H.}\ \bibnamefont
  {Busse}},\ }\bibfield  {title} {\bibinfo {title} {{Patterns of convection in
  spherical shells}},\ }\href {https://doi.org/10.1017/S0022112075002947}
  {\bibfield  {journal} {\bibinfo  {journal} {J. Fluid Mech.}\ }\textbf
  {\bibinfo {volume} {72}},\ \bibinfo {pages} {67} (\bibinfo {year}
  {1975})}\BibitemShut {NoStop}%
\bibitem [{\citenamefont {Weyl}(1952)}]{Weyl1952}%
  \BibitemOpen
  \bibfield  {author} {\bibinfo {author} {\bibfnamefont {H.}~\bibnamefont
  {Weyl}},\ }\href@noop {} {\emph {\bibinfo {title} {{Symmetry}}}}\ (\bibinfo
  {publisher} {Princeton University Press},\ \bibinfo {year}
  {1952})\BibitemShut {NoStop}%
\bibitem [{\citenamefont {Busse}\ and\ \citenamefont
  {Riahi}(1982)}]{Busse1982}%
  \BibitemOpen
  \bibfield  {author} {\bibinfo {author} {\bibfnamefont {F.~H.}\ \bibnamefont
  {Busse}}\ and\ \bibinfo {author} {\bibfnamefont {N.}~\bibnamefont {Riahi}},\
  }\bibfield  {title} {\bibinfo {title} {{Patterns of convection in spherical
  shells. Part 2}},\ }\href {https://doi.org/10.1017/S0022112082003061}
  {\bibfield  {journal} {\bibinfo  {journal} {J. Fluid Mech.}\ }\textbf
  {\bibinfo {volume} {123}},\ \bibinfo {pages} {283} (\bibinfo {year}
  {1982})}\BibitemShut {NoStop}%
\bibitem [{\citenamefont {Bercovici}\ \emph {et~al.}(1989)\citenamefont
  {Bercovici}, \citenamefont {Schubert}, \citenamefont {Glatzmaier},\ and\
  \citenamefont {Zebib}}]{Bercovici1989}%
  \BibitemOpen
  \bibfield  {author} {\bibinfo {author} {\bibfnamefont {D.}~\bibnamefont
  {Bercovici}}, \bibinfo {author} {\bibfnamefont {G.}~\bibnamefont {Schubert}},
  \bibinfo {author} {\bibfnamefont {G.~A.}\ \bibnamefont {Glatzmaier}},\ and\
  \bibinfo {author} {\bibfnamefont {A.}~\bibnamefont {Zebib}},\ }\bibfield
  {title} {\bibinfo {title} {{Three-Dimensional thermal convection in a
  spherical shell}},\ }\href {https://doi.org/10.1017/S0022112089002235}
  {\bibfield  {journal} {\bibinfo  {journal} {J. Fluid Mech.}\ }\textbf
  {\bibinfo {volume} {206}},\ \bibinfo {pages} {75} (\bibinfo {year}
  {1989})}\BibitemShut {NoStop}%
\bibitem [{\citenamefont {Futterer}\ \emph {et~al.}(2010)\citenamefont
  {Futterer}, \citenamefont {Egbers}, \citenamefont {Dahley}, \citenamefont
  {Koch},\ and\ \citenamefont {Jehring}}]{Futterer2010}%
  \BibitemOpen
  \bibfield  {author} {\bibinfo {author} {\bibfnamefont {B.}~\bibnamefont
  {Futterer}}, \bibinfo {author} {\bibfnamefont {C.}~\bibnamefont {Egbers}},
  \bibinfo {author} {\bibfnamefont {N.}~\bibnamefont {Dahley}}, \bibinfo
  {author} {\bibfnamefont {S.}~\bibnamefont {Koch}},\ and\ \bibinfo {author}
  {\bibfnamefont {L.}~\bibnamefont {Jehring}},\ }\bibfield  {title} {\bibinfo
  {title} {{First identification of sub- and supercritical convection patterns
  from 'GeoFlow', the geophysical flow simulation experiment integrated in
  Fluid Science Laboratory}},\ }\href
  {https://doi.org/10.1016/j.actaastro.2009.05.027} {\bibfield  {journal}
  {\bibinfo  {journal} {Acta Astronautica}\ }\textbf {\bibinfo {volume} {66}},\
  \bibinfo {pages} {193} (\bibinfo {year} {2010})}\BibitemShut {NoStop}%
\bibitem [{\citenamefont {Feudel}\ \emph {et~al.}(2011)\citenamefont {Feudel},
  \citenamefont {Bergemann}, \citenamefont {Tuckerman}, \citenamefont {Egbers},
  \citenamefont {Futterer}, \citenamefont {Gellert},\ and\ \citenamefont
  {Hollerbach}}]{Feudel2011}%
  \BibitemOpen
  \bibfield  {author} {\bibinfo {author} {\bibfnamefont {F.}~\bibnamefont
  {Feudel}}, \bibinfo {author} {\bibfnamefont {K.}~\bibnamefont {Bergemann}},
  \bibinfo {author} {\bibfnamefont {L.~S.}\ \bibnamefont {Tuckerman}}, \bibinfo
  {author} {\bibfnamefont {C.}~\bibnamefont {Egbers}}, \bibinfo {author}
  {\bibfnamefont {B.}~\bibnamefont {Futterer}}, \bibinfo {author}
  {\bibfnamefont {M.}~\bibnamefont {Gellert}},\ and\ \bibinfo {author}
  {\bibfnamefont {R.}~\bibnamefont {Hollerbach}},\ }\bibfield  {title}
  {\bibinfo {title} {{Convection patterns in a spherical fluid shell}},\ }\href
  {https://doi.org/10.1103/PhysRevE.83.046304} {\bibfield  {journal} {\bibinfo
  {journal} {Phys. Rev. E}\ }\textbf {\bibinfo {volume} {83}},\ \bibinfo
  {pages} {046304} (\bibinfo {year} {2011})}\BibitemShut {NoStop}%
\bibitem [{\citenamefont {Arrial}\ \emph {et~al.}(2014)\citenamefont {Arrial},
  \citenamefont {Flyer}, \citenamefont {Wright},\ and\ \citenamefont
  {Kellogg}}]{Arrial2014}%
  \BibitemOpen
  \bibfield  {author} {\bibinfo {author} {\bibfnamefont {P.~A.}\ \bibnamefont
  {Arrial}}, \bibinfo {author} {\bibfnamefont {N.}~\bibnamefont {Flyer}},
  \bibinfo {author} {\bibfnamefont {G.~B.}\ \bibnamefont {Wright}},\ and\
  \bibinfo {author} {\bibfnamefont {L.~H.}\ \bibnamefont {Kellogg}},\
  }\bibfield  {title} {\bibinfo {title} {{On the sensitivity of 3-D thermal
  convection codes to numerical discretization: A model intercomparison}},\
  }\href {https://doi.org/10.5194/gmd-7-2065-2014} {\bibfield  {journal}
  {\bibinfo  {journal} {Geosci. Model Dev.}\ }\textbf {\bibinfo {volume} {7}},\
  \bibinfo {pages} {2065} (\bibinfo {year} {2014})}\BibitemShut {NoStop}%
\bibitem [{\citenamefont {Matthews}(2003)}]{Matthews2003}%
  \BibitemOpen
  \bibfield  {author} {\bibinfo {author} {\bibfnamefont {P.~C.}\ \bibnamefont
  {Matthews}},\ }\bibfield  {title} {\bibinfo {title} {{Pattern formation on a
  sphere}},\ }\href {https://doi.org/10.1103/PhysRevE.67.036206} {\bibfield
  {journal} {\bibinfo  {journal} {Phys. Rev. E}\ }\textbf {\bibinfo {volume}
  {67}},\ \bibinfo {pages} {036206} (\bibinfo {year} {2003})}\BibitemShut
  {NoStop}%
\bibitem [{\citenamefont {Hoyle}(2006)}]{Hoyle2006}%
  \BibitemOpen
  \bibfield  {author} {\bibinfo {author} {\bibfnamefont {R.}~\bibnamefont
  {Hoyle}},\ }\href {https://doi.org/10.1017/CBO9780511616051} {\emph {\bibinfo
  {title} {Pattern Formation: An Introduction to Methods}}}\ (\bibinfo
  {publisher} {Cambridge University Press},\ \bibinfo {year} {2006})\ pp.\
  \bibinfo {pages} {1--422}\BibitemShut {NoStop}%
\bibitem [{\citenamefont {Chossat}\ and\ \citenamefont
  {Lauterbach}(2000)}]{Chossat2000}%
  \BibitemOpen
  \bibfield  {author} {\bibinfo {author} {\bibfnamefont {P.}~\bibnamefont
  {Chossat}}\ and\ \bibinfo {author} {\bibfnamefont {R.}~\bibnamefont
  {Lauterbach}},\ }\href {https://doi.org/10.1142/4062} {\emph {\bibinfo
  {title} {{Methods in Equivariant Bifurcations and Dynamical Systems}}}},\
  Advanced Series in Nonlinear Dynamics\ (\bibinfo  {publisher} {WORLD
  SCIENTIFIC},\ \bibinfo {year} {2000})\BibitemShut {NoStop}%
\bibitem [{\citenamefont {Sigrist}\ and\ \citenamefont
  {Matthews}(2011)}]{Sigrist2011}%
  \BibitemOpen
  \bibfield  {author} {\bibinfo {author} {\bibfnamefont {R.}~\bibnamefont
  {Sigrist}}\ and\ \bibinfo {author} {\bibfnamefont {P.}~\bibnamefont
  {Matthews}},\ }\bibfield  {title} {\bibinfo {title} {{Symmetric spiral
  patterns on spheres}},\ }\href {https://doi.org/10.1137/100806692} {\bibfield
   {journal} {\bibinfo  {journal} {SIAM J. Appl. Dyn. Syst.,}\ }\textbf
  {\bibinfo {volume} {10}},\ \bibinfo {pages} {1177} (\bibinfo {year}
  {2011})}\BibitemShut {NoStop}%
\bibitem [{\citenamefont {Turing}(1952)}]{Turing1952}%
  \BibitemOpen
  \bibfield  {author} {\bibinfo {author} {\bibfnamefont {A.~M.}\ \bibnamefont
  {Turing}},\ }\bibfield  {title} {\bibinfo {title} {{The Chemical Basis of
  Morphogenesis}},\ }\href {http://www.jstor.org/about/terms.html.} {\bibfield
  {journal} {\bibinfo  {journal} {Philos. Trans. R. Soc. Lond., Ser. B}\
  }\textbf {\bibinfo {volume} {237}},\ \bibinfo {pages} {37} (\bibinfo {year}
  {1952})}\BibitemShut {NoStop}%
\bibitem [{\citenamefont {Varea}\ \emph {et~al.}(1999)\citenamefont {Varea},
  \citenamefont {Arag{\'{o}}n},\ and\ \citenamefont {Barrio}}]{Varea1999}%
  \BibitemOpen
  \bibfield  {author} {\bibinfo {author} {\bibfnamefont {C.}~\bibnamefont
  {Varea}}, \bibinfo {author} {\bibfnamefont {J.~L.}\ \bibnamefont
  {Arag{\'{o}}n}},\ and\ \bibinfo {author} {\bibfnamefont {R.~A.}\ \bibnamefont
  {Barrio}},\ }\bibfield  {title} {\bibinfo {title} {{Turing patterns on a
  sphere}},\ }\href@noop {} {\bibfield  {journal} {\bibinfo  {journal} {Phys.
  Rev. E}\ }\textbf {\bibinfo {volume} {60}},\ \bibinfo {pages} {4588}
  (\bibinfo {year} {1999})}\BibitemShut {NoStop}%
\bibitem [{\citenamefont {S{\'{a}}nchez-Gardu{\~{n}}o}\ \emph
  {et~al.}(2019)\citenamefont {S{\'{a}}nchez-Gardu{\~{n}}o}, \citenamefont
  {Krause}, \citenamefont {Castillo},\ and\ \citenamefont
  {Padilla}}]{Sanchez2019}%
  \BibitemOpen
  \bibfield  {author} {\bibinfo {author} {\bibfnamefont {F.}~\bibnamefont
  {S{\'{a}}nchez-Gardu{\~{n}}o}}, \bibinfo {author} {\bibfnamefont {A.~L.}\
  \bibnamefont {Krause}}, \bibinfo {author} {\bibfnamefont {J.~A.}\
  \bibnamefont {Castillo}},\ and\ \bibinfo {author} {\bibfnamefont
  {P.}~\bibnamefont {Padilla}},\ }\bibfield  {title} {\bibinfo {title}
  {{Turing-Hopf patterns on growing domains: The torus and the sphere}},\
  }\href {https://doi.org/10.1016/j.jtbi.2018.09.028} {\bibfield  {journal}
  {\bibinfo  {journal} {J. Theor. Biol.}\ }\textbf {\bibinfo {volume} {481}},\
  \bibinfo {pages} {136} (\bibinfo {year} {2019})}\BibitemShut {NoStop}%
\bibitem [{\citenamefont {Zhang}\ \emph {et~al.}(2002)\citenamefont {Zhang},
  \citenamefont {Liao},\ and\ \citenamefont {Zhang}}]{Zhang2002a}%
  \BibitemOpen
  \bibfield  {author} {\bibinfo {author} {\bibfnamefont {P.}~\bibnamefont
  {Zhang}}, \bibinfo {author} {\bibfnamefont {X.}~\bibnamefont {Liao}},\ and\
  \bibinfo {author} {\bibfnamefont {K.}~\bibnamefont {Zhang}},\ }\bibfield
  {title} {\bibinfo {title} {{Patterns in spherical Rayleigh-B{\'{e}}nard
  convection: A giant spiral roll and its dislocations}},\ }\href
  {https://doi.org/10.1103/PhysRevE.66.055203} {\bibfield  {journal} {\bibinfo
  {journal} {Phys. Rev. E}\ }\textbf {\bibinfo {volume} {66}},\ \bibinfo
  {pages} {055203(R)} (\bibinfo {year} {2002})}\BibitemShut {NoStop}%
\bibitem [{\citenamefont {Itano}\ \emph {et~al.}(2015)\citenamefont {Itano},
  \citenamefont {Ninomiya}, \citenamefont {Konno},\ and\ \citenamefont
  {Sugihara-Seki}}]{Itano2015}%
  \BibitemOpen
  \bibfield  {author} {\bibinfo {author} {\bibfnamefont {T.}~\bibnamefont
  {Itano}}, \bibinfo {author} {\bibfnamefont {T.}~\bibnamefont {Ninomiya}},
  \bibinfo {author} {\bibfnamefont {K.}~\bibnamefont {Konno}},\ and\ \bibinfo
  {author} {\bibfnamefont {M.}~\bibnamefont {Sugihara-Seki}},\ }\bibfield
  {title} {\bibinfo {title} {{Spiral roll state in heat convection between
  nonrotating concentric double spherical boundaries}},\ }\href
  {https://doi.org/10.7566/JPSJ.84.103401} {\bibfield  {journal} {\bibinfo
  {journal} {J. Phys. Soc. Jpn.}\ }\textbf {\bibinfo {volume} {84}},\ \bibinfo
  {pages} {103401} (\bibinfo {year} {2015})}\BibitemShut {NoStop}%
\bibitem [{\citenamefont {Morris}\ \emph {et~al.}(1993)\citenamefont {Morris},
  \citenamefont {Bodenschatz}, \citenamefont {Cannell},\ and\ \citenamefont
  {Ahlers}}]{Morris1993}%
  \BibitemOpen
  \bibfield  {author} {\bibinfo {author} {\bibfnamefont {S.~W.}\ \bibnamefont
  {Morris}}, \bibinfo {author} {\bibfnamefont {E.}~\bibnamefont {Bodenschatz}},
  \bibinfo {author} {\bibfnamefont {D.~S.}\ \bibnamefont {Cannell}},\ and\
  \bibinfo {author} {\bibfnamefont {G.}~\bibnamefont {Ahlers}},\ }\bibfield
  {title} {\bibinfo {title} {{Spiral Defect Chaos in Large Aspect Ratio
  Rayleigh-Benard Convection}},\ }\href@noop {} {\bibfield  {journal} {\bibinfo
   {journal} {Phys. Rev. Lett.}\ }\textbf {\bibinfo {volume} {71}},\ \bibinfo
  {pages} {2026} (\bibinfo {year} {1993})}\BibitemShut {NoStop}%
\bibitem [{\citenamefont {Xi}\ \emph {et~al.}(1993)\citenamefont {Xi},
  \citenamefont {Gunton},\ and\ \citenamefont {Vinals}}]{Xi1993}%
  \BibitemOpen
  \bibfield  {author} {\bibinfo {author} {\bibfnamefont {H.-W.}\ \bibnamefont
  {Xi}}, \bibinfo {author} {\bibfnamefont {J.~D.}\ \bibnamefont {Gunton}},\
  and\ \bibinfo {author} {\bibfnamefont {J.}~\bibnamefont {Vinals}},\
  }\bibfield  {title} {\bibinfo {title} {{Spiral Defect Chaos in a Model of
  Rayleigh-Benard Convection}},\ }\href@noop {} {\bibfield  {journal} {\bibinfo
   {journal} {Phys. Rev. Lett.}\ }\textbf {\bibinfo {volume} {71}},\ \bibinfo
  {pages} {2030} (\bibinfo {year} {1993})}\BibitemShut {NoStop}%
\bibitem [{\citenamefont {Liu}\ and\ \citenamefont {Ahlers}(1996)}]{Liu1996}%
  \BibitemOpen
  \bibfield  {author} {\bibinfo {author} {\bibfnamefont {J.}~\bibnamefont
  {Liu}}\ and\ \bibinfo {author} {\bibfnamefont {G.}~\bibnamefont {Ahlers}},\
  }\bibfield  {title} {\bibinfo {title} {{Spiral-Defect Chaos in
  Rayleigh-B{\'{e}}nard Convection with Small Prandtl Numbers}},\ }\href@noop
  {} {\bibfield  {journal} {\bibinfo  {journal} {Phys. Rev. Lett.}\ }\textbf
  {\bibinfo {volume} {77}},\ \bibinfo {pages} {3126} (\bibinfo {year}
  {1996})}\BibitemShut {NoStop}%
\bibitem [{\citenamefont {Ecke}\ \emph {et~al.}(1995)\citenamefont {Ecke},
  \citenamefont {Hu}, \citenamefont {Mainieri},\ and\ \citenamefont
  {Ahlers}}]{Ecke1995}%
  \BibitemOpen
  \bibfield  {author} {\bibinfo {author} {\bibfnamefont {R.~E.}\ \bibnamefont
  {Ecke}}, \bibinfo {author} {\bibfnamefont {Y.}~\bibnamefont {Hu}}, \bibinfo
  {author} {\bibfnamefont {R.}~\bibnamefont {Mainieri}},\ and\ \bibinfo
  {author} {\bibfnamefont {G.}~\bibnamefont {Ahlers}},\ }\bibfield  {title}
  {\bibinfo {title} {{Excitation of Spirals and Chiral Symmetry Breaking in
  Rayleigh-B{\'{e}}nard Convection}},\ }\href@noop {} {\bibfield  {journal}
  {\bibinfo  {journal} {Science}\ }\textbf {\bibinfo {volume} {269}},\ \bibinfo
  {pages} {1704} (\bibinfo {year} {1995})}\BibitemShut {NoStop}%
\bibitem [{\citenamefont {Gilman}\ and\ \citenamefont
  {Glatzmaier}(1981)}]{Gilman1981}%
  \BibitemOpen
  \bibfield  {author} {\bibinfo {author} {\bibfnamefont {P.}~\bibnamefont
  {Gilman}}\ and\ \bibinfo {author} {\bibfnamefont {G.~A.}\ \bibnamefont
  {Glatzmaier}},\ }\bibfield  {title} {\bibinfo {title} {{Compressible
  convection in a rotating spherical shell. I}},\ }\href@noop {} {\bibfield
  {journal} {\bibinfo  {journal} {Astrophys. J. Suppl.}\ }\textbf {\bibinfo
  {volume} {45}},\ \bibinfo {pages} {335} (\bibinfo {year} {1981})}\BibitemShut
  {NoStop}%
\bibitem [{\citenamefont {Spiegel}(1965)}]{Spiegel1965}%
  \BibitemOpen
  \bibfield  {author} {\bibinfo {author} {\bibfnamefont {E.~A.}\ \bibnamefont
  {Spiegel}},\ }\bibfield  {title} {\bibinfo {title} {{Convective Instability
  in a Compressible Atmosphere. I}},\ }\href@noop {} {\bibfield  {journal}
  {\bibinfo  {journal} {Astrophys. J.}\ }\textbf {\bibinfo {volume} {141}},\
  \bibinfo {pages} {1068} (\bibinfo {year} {1965})}\BibitemShut {NoStop}%
\bibitem [{\citenamefont {Kageyama}\ and\ \citenamefont
  {Sato}(2004)}]{Kageyama2004}%
  \BibitemOpen
  \bibfield  {author} {\bibinfo {author} {\bibfnamefont {A.}~\bibnamefont
  {Kageyama}}\ and\ \bibinfo {author} {\bibfnamefont {T.}~\bibnamefont
  {Sato}},\ }\bibfield  {title} {\bibinfo {title} {{"Yin-Yang grid": An overset
  grid in spherical geometry}},\ }\href {https://doi.org/10.1029/2004GC000734}
  {\bibfield  {journal} {\bibinfo  {journal} {Geochem. Geophys. Geosyst.}\
  }\textbf {\bibinfo {volume} {5}},\ \bibinfo {pages} {Q09005} (\bibinfo {year}
  {2004})}\BibitemShut {NoStop}%
\bibitem [{\citenamefont {Ohno}\ and\ \citenamefont
  {Kageyama}(2021)}]{Ohno2021b}%
  \BibitemOpen
  \bibfield  {author} {\bibinfo {author} {\bibfnamefont {N.}~\bibnamefont
  {Ohno}}\ and\ \bibinfo {author} {\bibfnamefont {A.}~\bibnamefont
  {Kageyama}},\ }\bibfield  {title} {\bibinfo {title} {{In-situ visualization
  library for Yin-Yang grid simulations}},\ }\href@noop {} {\bibfield
  {journal} {\bibinfo  {journal} {Earth, Planets and Space}\ }\textbf {\bibinfo
  {volume} {73}},\ \bibinfo {pages} {158} (\bibinfo {year} {2021})}\BibitemShut
  {NoStop}%
\bibitem [{\citenamefont {Ohno}()}]{Ohno}%
  \BibitemOpen
  \bibfield  {author} {\bibinfo {author} {\bibfnamefont {N.}~\bibnamefont
  {Ohno}},\ }\href@noop {} {\bibinfo {title}
  {https://vizlab.sakura.ne.jp/en/vismo.en.html}}\BibitemShut {NoStop}%
\bibitem [{\citenamefont {Kageyama}\ and\ \citenamefont
  {Sakamoto}(2020)}]{Kageyama2020}%
  \BibitemOpen
  \bibfield  {author} {\bibinfo {author} {\bibfnamefont {A.}~\bibnamefont
  {Kageyama}}\ and\ \bibinfo {author} {\bibfnamefont {N.}~\bibnamefont
  {Sakamoto}},\ }\bibfield  {title} {\bibinfo {title} {{4D street view: a
  video-based visualization method}},\ }\href
  {https://doi.org/10.7717/peerj-cs.305} {\bibfield  {journal} {\bibinfo
  {journal} {PeerJ Comput. Sci.}\ }\textbf {\bibinfo {volume} {6}},\ \bibinfo
  {pages} {e305} (\bibinfo {year} {2020})}\BibitemShut {NoStop}%
\bibitem [{\citenamefont {Chandrasekhar}(1981)}]{Chandrasekhar1981a}%
  \BibitemOpen
  \bibfield  {author} {\bibinfo {author} {\bibfnamefont {S.~S.}\ \bibnamefont
  {Chandrasekhar}},\ }\href@noop {} {\emph {\bibinfo {title} {{Hydrodynamic and
  hydromagnetic stability}}}}\ (\bibinfo  {publisher} {Dover Publications},\
  \bibinfo {year} {1981})\BibitemShut {NoStop}%
\bibitem [{\citenamefont {Chen}\ and\ \citenamefont
  {Whitehead}(1968)}]{Chen1968}%
  \BibitemOpen
  \bibfield  {author} {\bibinfo {author} {\bibfnamefont {M.~M.}\ \bibnamefont
  {Chen}}\ and\ \bibinfo {author} {\bibfnamefont {J.~A.}\ \bibnamefont
  {Whitehead}},\ }\bibfield  {title} {\bibinfo {title} {{Evolution of
  two-dimensional periodic Rayleigh convection cells of arbitrary
  wave-numbers}},\ }\href {https://doi.org/10.1017/S0022112068000017}
  {\bibfield  {journal} {\bibinfo  {journal} {J. Fluid Mech.}\ }\textbf
  {\bibinfo {volume} {31}},\ \bibinfo {pages} {1} (\bibinfo {year}
  {1968})}\BibitemShut {NoStop}%
\bibitem [{\citenamefont {Busse}\ and\ \citenamefont
  {Whitehead}(1974)}]{Busse1974}%
  \BibitemOpen
  \bibfield  {author} {\bibinfo {author} {\bibfnamefont {F.~H.}\ \bibnamefont
  {Busse}}\ and\ \bibinfo {author} {\bibfnamefont {J.~A.}\ \bibnamefont
  {Whitehead}},\ }\bibfield  {title} {\bibinfo {title} {{Oscillatory and
  collective instabilities in large Prandtl number convection}},\ }\href
  {https://doi.org/10.1017/S0022112074000061} {\bibfield  {journal} {\bibinfo
  {journal} {J. Fluid Mech.}\ }\textbf {\bibinfo {volume} {66}},\ \bibinfo
  {pages} {67} (\bibinfo {year} {1974})}\BibitemShut {NoStop}%
\bibitem [{\citenamefont {Cromwell}(1997)}]{Cromwell1997}%
  \BibitemOpen
  \bibfield  {author} {\bibinfo {author} {\bibfnamefont {P.}~\bibnamefont
  {Cromwell}},\ }\href
  {https://www.cambridge.org/jp/academic/subjects/mathematics/geometry-and-topology/polyhedra?format=PB&isbn=9780521664059}
  {\emph {\bibinfo {title} {{Polyhedra}}}}\ (\bibinfo  {publisher} {Cambridge
  University Press},\ \bibinfo {year} {1997})\BibitemShut {NoStop}%
\bibitem [{\citenamefont {Kageyama}()}]{KageyamaURL}%
  \BibitemOpen
  \bibfield  {author} {\bibinfo {author} {\bibfnamefont {A.}~\bibnamefont
  {Kageyama}},\ }\href@noop {} {\bibinfo {title}
  {https://github.com/akageyama/spirals-on-dodecahedron}}\BibitemShut {NoStop}%
\bibitem [{\citenamefont {Bodenschatz}\ \emph {et~al.}(1991)\citenamefont
  {Bodenschatz}, \citenamefont {de~Bruyn}, \citenamefont {Ahlers},\ and\
  \citenamefont {Cannell}}]{Bodenschatz1991a}%
  \BibitemOpen
  \bibfield  {author} {\bibinfo {author} {\bibfnamefont {E.}~\bibnamefont
  {Bodenschatz}}, \bibinfo {author} {\bibfnamefont {J.~R.}\ \bibnamefont
  {de~Bruyn}}, \bibinfo {author} {\bibfnamefont {G.}~\bibnamefont {Ahlers}},\
  and\ \bibinfo {author} {\bibfnamefont {D.~S.}\ \bibnamefont {Cannell}},\
  }\bibfield  {title} {\bibinfo {title} {{Transitions between Patterns in
  Thermal Convection}},\ }\href@noop {} {\bibfield  {journal} {\bibinfo
  {journal} {Phys. Rev. Lett.}\ }\textbf {\bibinfo {volume} {67}},\ \bibinfo
  {pages} {3078} (\bibinfo {year} {1991})}\BibitemShut {NoStop}%
\bibitem [{\citenamefont {Vitral}\ \emph {et~al.}(2020)\citenamefont {Vitral},
  \citenamefont {Mukherjee}, \citenamefont {Leo}, \citenamefont {Vinals},
  \citenamefont {Paul},\ and\ \citenamefont {Huang}}]{Vitral2020}%
  \BibitemOpen
  \bibfield  {author} {\bibinfo {author} {\bibfnamefont {E.}~\bibnamefont
  {Vitral}}, \bibinfo {author} {\bibfnamefont {S.}~\bibnamefont {Mukherjee}},
  \bibinfo {author} {\bibfnamefont {P.~H.}\ \bibnamefont {Leo}}, \bibinfo
  {author} {\bibfnamefont {J.}~\bibnamefont {Vinals}}, \bibinfo {author}
  {\bibfnamefont {M.~R.}\ \bibnamefont {Paul}},\ and\ \bibinfo {author}
  {\bibfnamefont {Z.-F.}\ \bibnamefont {Huang}},\ }\bibfield  {title} {\bibinfo
  {title} {{Spiral defect chaos in Rayleigh-Benard convection: Asymptotic and
  numerical studies of azimuthal flows induced by rotating spirals}},\ }\href
  {https://doi.org/10.1103/PhysRevFluids.5.093501} {\bibfield  {journal}
  {\bibinfo  {journal} {Phys. Rev. Fluids}\ }\textbf {\bibinfo {volume} {5}},\
  \bibinfo {pages} {093501} (\bibinfo {year} {2020})}\BibitemShut {NoStop}%
\end{thebibliography}
%-------
% paste multiple_spirals.bbl below
%-------


\end{document}